\documentclass[preprint,amsmath,amssymb,prb,]{revtex4}

\usepackage{graphicx}
\usepackage{dcolumn}
\usepackage{bm}

\begin{document}

\title{Incorporating quasiparticle and excitonic properties into material discovery}

\author{Tathagata Biswas}
\author{Arunima K. Singh}%
 \email{arunimasingh@asu.edu}
\affiliation{Department of Physics, Arizona State University, Tempe, Arizona, 85281, USA}

\date{\today}

\begin{abstract}
	In recent years, GW-BSE has been proven to be extremely successful in studying
	the quasiparticle (QP) bandstructures and excitonic effects in the
	optical properties of materials. However, the massive computational
	cost associated with such calculations restricts their applicability in
	high-throughput material discovery studies. Recently, we developed a
	Python workflow package, $py$GWBSE, to perform high-throughput GW-BSE
	simulations. In this work, using $py$GWBSE we create a database of
	various QP properties and excitonic properties of over 350 chemically
	and structurally diverse materials. Despite the relatively small size
	of the dataset, we obtain highly accurate supervised machine learning
	(ML) models via the dataset. The models predict the quasiparticle gap
	with an RMSE of 0.36 eV, exciton binding energies of materials with an
	RMSE of 0.29 eV, and classify materials as high or low excitonic
	binding energy materials with classification accuracy of 90\%. We
	exemplify the application of these ML models in the discovery of 159
	visible-light and 203 ultraviolet-light photoabsorber materials
	utilizing the Materials Project database. 
\end{abstract}
\maketitle

\section{Introduction}

Light-matter interaction is the fundamental physical phenomenon behind a wide variety of existing high-impact applications such as photovoltaics, photocatalysis, medical diagnostics, scientific instrumentation, imaging (e.g., infrared imagers), and sensing (e.g., for light detection) along with potential future quantum devices \cite{Borys2022,biswas2021excitonic,hasan2016antimony}. New opportunities to optimize the performance of these existing optical devices and to pave the road for emerging fields can be realized by the discovery and design of novel functional materials. 

The Materials Genome Initiative (MGI) \cite{de2014materials, MGI} was proposed in 2011 to enable the discovery, manufacturing, and deployment of advanced materials twice as fast and at a fraction of the cost compared to traditional methods. To achieve the MGI objectives one of the key strategies adopted was to harness the power of data and computational tools jointly with experimental investigations \cite{de2014materials, MGI}. Since then a new paradigm for accelerated materials discovery has emerged by designing new compounds in silico using first-principles calculations and then performing experiments on the computationally designed candidates \cite{shinde2017discovery,zhou2018rutile,singh2019robust}. Several open-source databases have been developed to aid the accelerated material discovery goal such as the Materials Project \cite{jain2013commentary}, Aflowlib \cite{curtarolo2012aflowlib}, C2DB \cite{haastrup2018computational}, ESP \cite{ortiz2009data}, NoMaD\cite{nomad}, OQMD \cite{kirklin2015open} etc. The availability of such large data has opened up an emerging paradigm, the application of machine learning (ML) and other data science methods for material discovery, thus, making material discovery essentially a big-data problem \cite{yao2023machine,saal2020machine,liu2017materials,pyzer2022accelerating,singh2022data,torrisi2020two}. ML accelerated material discovery has made a revolutionary impact in applications ranging from organic and solid-state LEDs, batteries, ferroelectric, high-k dielectric, hydrogen storage, high-entropy alloys, and thermoplastics to shape memory alloys \cite{saal2020machine,rao2022machine,singh2022data}.    
 
The fundamental challenge of applying this well-established material discovery paradigm of combining first-principles computations and data science methods to applications where light-matter interaction is a key phenomenon is the unavailability of large datasets that are accurate enough in comparison to experimental observations. While some first-principles methods such as GW-BSE (Bethe-Salpeter equation) formalism for simulating excitonic effects can produce optical properties with sufficient accuracy they are computationally very expensive. Thus it is not surprising that the largest database of GW-BSE computed absorption spectra is that of $\sim$ 300 spectra of two-dimensional (2D) materials \cite{haastrup2018computational}. 

The unavailability of large-scale data of first principles computed excited state properties can also be partially attributed to the lack of open-source computational tools to perform such high-throughput computations. Recently, the authors have developed $py$GWBSE \cite{biswas2023py}, a python workflow package that enables automated high-throughput GW-BSE simulations and made it available through open-source licensing. 

In this study, we demonstrate the applicability of ML models in predicting accurate excited state properties such as quasi-particle gap (QPG) and exciton binding energy (EBE). To accomplish this goal, we generated the largest database of first-principles computed QP and excitonic properties of bulk materials computed and curated using $py$GWBSE. This database contains static dielectric constants, effective masses, QP bandstructure, absorption spectra, and several other excited state properties such as EBE, integrated absorption coefficients, etc. of more than 350 bulk materials, and new materials are being added to the database continuously. We find that among the various ML regression algorithms, Random Forest Regression has the best performance for predicting QPG within an RMSE of 0.36 and EBE within an RMSE of 0.29 eV. We also find ML models that accurately classify materials to have integrated absorption coefficient (IAC),~\cite{biswas2021excitonic} anisotropy in absorption coefficient (AAC),~\cite{biswas2021excitonic} and excitonic binding energies suited for a good photoabsorber with an accuracy of $\sim$ 90 \%. Lastly, we apply the ML models developed in this work to identify promising materials that can absorb visible and ultraviolet (UV) radiation for photovoltaic or photocatalytic applications based on their QP and excitonic properties from a list of $\sim$ 7000 materials for which only the ground state properties are available in the Materials Project database. 


\section{Results and Discussions}
\label{results}

\begin{figure}
\centering
  \includegraphics[width=\linewidth]{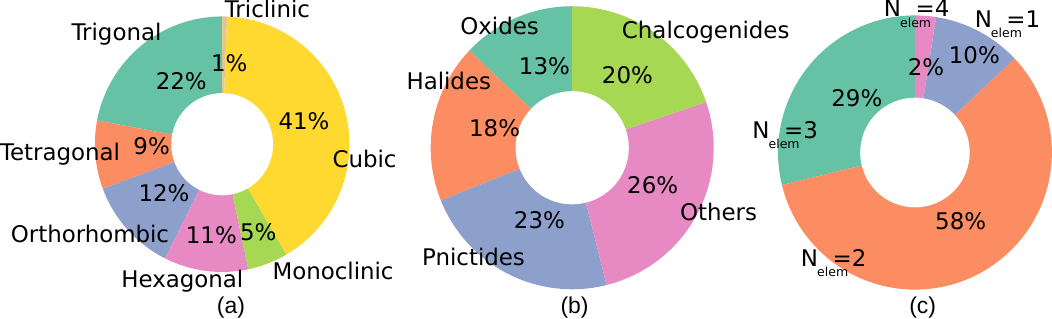}
\caption{A classification of the bulk materials' dataset considered in this work in terms of (a) the crystal system, (b) chemical composition, and (c) the number of constituent elements. The figure shows that the materials in the dataset are very diverse.}
\label{fig1}
\end{figure}

As we intend to develop and apply ML models for predicting the QP and excitonic properties of a wide variety of materials, we first demonstrate the diversity of the materials within our GW-BSE computed dataset. In Fig. \ref{fig1} we classify all the materials in the dataset according to their chemical composition, crystal systems, and the number of constituent atomic species. As one can see from Fig. \ref{fig1}(a), our materials dataset has a significant fraction of materials from all the seven crystal systems, except the triclinic phase,  which constitutes less than 1 $\%$ of our training set. Fig. \ref{fig1}(b) shows that our training set of materials consists of almost equal percentages of oxides, pnictides, halides, and chalcogenides. In addition, 26 $\%$ of the materials in the training set don't fall in any of the aforementioned chemical groups. Given the diversity in the chemical compositions and the crystal symmetries of the materials in our dataset, we can expect that the ML model trained with the data can be used for a wide variety of materials. Moreover, we also look at the number of unique chemical species in the selected materials in Fig \ref{fig1}(c) which shows that most of the materials in our dataset are binary (58 $\%$) and ternary (29 $\%$) compounds, which ensures that our ML model captures diverse chemical properties of multi-element compounds. 

The diversity of our GW-BSE computed database is unique in terms of all the aforementioned characteristics. The only other database that hosts GW-BSE computed properties is that of Hasstrup et al.\cite{haastrup2018computational} which is limited to two-dimensional (2D) materials and therefore restrictive in terms of crystal systems and chemical compositions.             

In the following two sections, sections \ref{qpg_pred} and \ref{ebe_pred}, we present the accuracy of various ML algorithms in predicting the QP and excitonic properties of materials. In particular, we focus on the QPG and the EBE of materials. We also discuss the most important features that were used in these models and their physical significance. In section \ref{classification} we present ML models for classifying materials as high or low excitonic binding energy materials. Finally, section \ref{screening} demonstrates how the ML models developed in this work allow the discovery of visible-light and UV-light photoabsorber materials by utilizing existing materials database without the need for any explicit GW-BSE simulations. 

\subsection{QP gap prediction}
\label{qpg_pred}

\begin{figure}
\centering
\includegraphics[width=\linewidth]{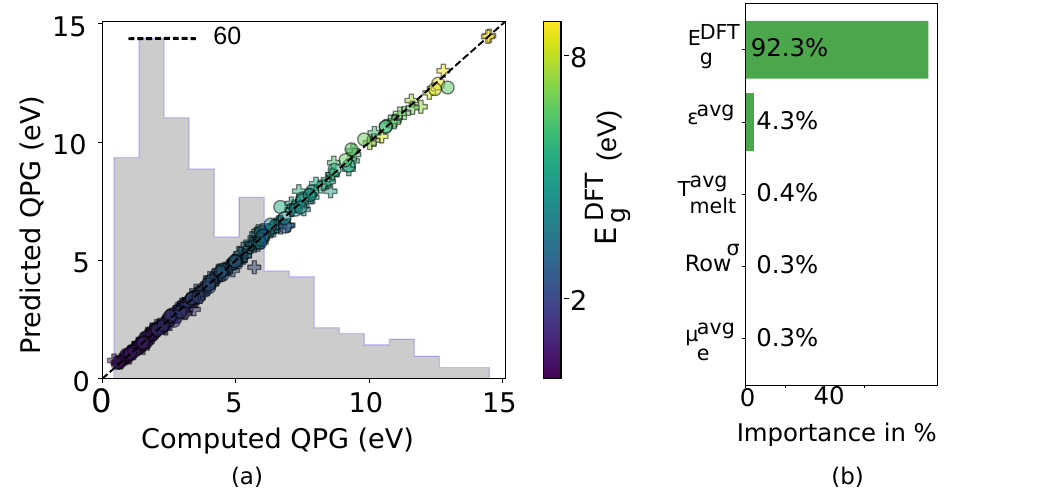}
\caption{ (a) Random Forest Regression model predicted QPG plotted against GW computed QPG. The training and test set data points are shown as $+$ symbols and circle symbols, respectively.  The distribution of the QPGs corresponding to the entire dataset is also shown in the background as a bar plot. To show the scale used for the histogram we have shown the highest value as reference in the figure. (b) The five most important features used to predict the QPG along with their percentage importance are shown as a bar plot. See section II of the supporting information for a detailed description of the features used in this study. The training and test set data points in (a) are colored based on the DFT computed bandgaps, E$_g^{DFT}$, which is the most important feature in the prediction. The E$_g^{DFT}$ is denoted by the color bar.} 
\label{fig2}
\end{figure}

The bar plot in Fig. \ref{fig2} (a) shows the distribution of the QP gaps of the 314 materials' dataset.  One can see that the materials have a broad range of QP gaps (0--15 eV). 

For the ML prediction of the QPG, we tested four regression methods-- the kernel ridge regression (KRR), the random forest regression (RF), the support vector machine (SVM), and the multi-layer perceptron (MLP) \cite{bonaccorso2017machine, mohammed2016machine} methods. A 10-fold cross-validation was employed to ensure randomness in the training and test datasets. Note that while our dataset consists of GW-BSE calculations of nearly 350 materials about 35 of them were not included in the ML model development as they had unphysical values for some of the features considered in this study. The model obtained from the RF method performed best for the QP bandgap prediction with an R$^2$ score of 0.98 and RMSE of only 0.36 eV. Fig. \ref{fig2} (a) compares the RF model predicted and GW computed QP gaps where the training set is shown with the `$+$` symbols and the test set via the circle symbols. The model obtained from the MLP also led to a very similar R$^2$ score as the one obtained from the RF method. Table S1 in the Supporting Information compares the performance of the models obtained from all four methods. 

It is noteworthy that most of the previous studies of ML-based bandgap predictions have been limited to a particular material class such as studies of MXenes by Rajan et. al. \cite{Rajan2018} or perovskites by Pilania et. al \cite{pilania2016machine}. In contrast, in this work, we have a dataset that includes materials without any such restrictions on chemical compositions or materials classes. In spite of working with a more diverse set of materials, our RF model is capable of similar/better accuracy as the earlier studies \cite{Rajan2018, pilania2016machine}. 

 Fig. \ref{fig2} (b) shows the most essential features for the QP gap predictions. The number of features used and their importance in the Random Forest algorithm is computed by calculating the Gini importance (see methods section \ref{ml-methods} for more details). As one might expect, Fig. \ref{fig2} (b) shows that the DFT computed bandgap is the most important feature in the QP gap prediction with 92~\% importance. Three of the 5 features shown in Fig. \ref{fig2} (b)
have very low total importance, less than 1~\%. We examine the relevance of these features by comparing the RMSE values of RF models that include 1 to 9 of the most important features.  An RF model obtained by including just the most important feature, i.e. the DFT gap, gives quite a large RMSE of 0.72 eV. Fig. S1 in the supporting information shows that the RMSE values for the QP gap prediction decrease from 0.44 eV to 0.36 eV when the number of features included in the model increases from 2 to 5, and thereafter it remains almost constant for up to 9 features.  Thus, not only is the second most important feature, i.e. mean dielectric constant (mean of $\epsilon_{x}$, $\epsilon_{y}$ and $\epsilon_{z}$), important for accurate QP gap predictions, the three other features are also needed despite their low, $<$ 1~\%, contribution to the total feature importance.

\subsection{EBE prediction}
\label{ebe_pred}

\begin{figure}
\centering
\includegraphics[width=\linewidth]{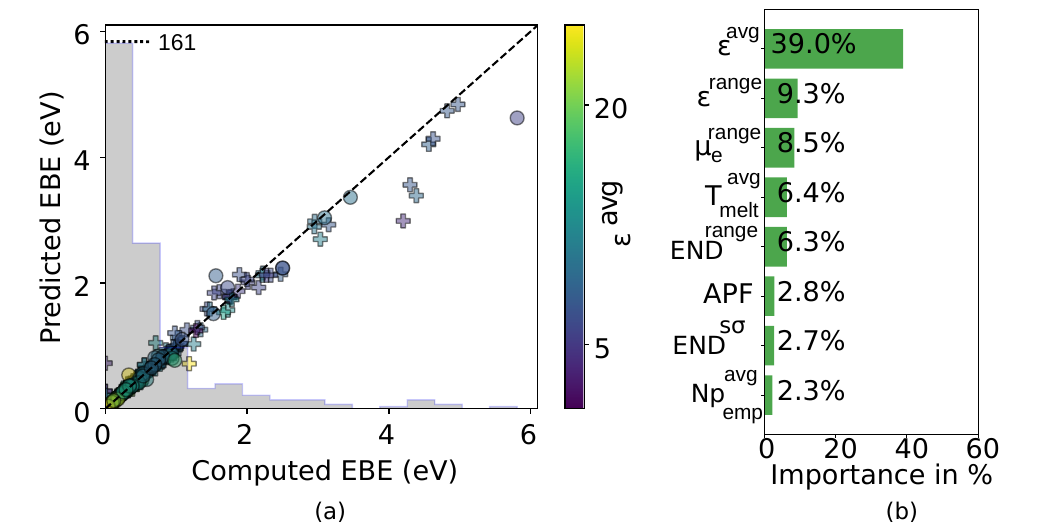}
\caption{(a) RF model predicted EBE plotted against GW-BSE computed EBE in the training ($+$ symbols) and test set (circle symbols). The distribution of EBEs corresponding to the entire dataset is also shown in the background as a bar plot. To show the scale used for the histogram we have shown the highest value as reference in the figure. (b) The eight most important features according to their Gini Importance and their \% importances are shown as a bar plot. See section II of the supporting information for a detailed description of all the features used in this study. The training and test set data points in (a) are colored based on the average dielectric constant, $\epsilon^\mathrm{avg}$, which is the most important feature in the EBE prediction. The $\epsilon^\mathrm{avg}$ value is denoted by the color bar.}
\label{fig3}
\end{figure}

Fig. \ref{fig3} (a) shows the distribution of the EBE of the materials in our dataset as a bar plot (shaded region in the background). We note that the dataset consists of materials with EBE in a very wide range, up to 6 eV. However, the majority of the materials (over 80~$\%$) have EBE < 1 eV. Fig. \ref{fig3} (a) also shows the RF model predicted EBE plotted against computed EBE for all the materials in our dataset. The RF model predicts the EBE of materials with an R$^2$ score of 0.86 and an RMSE of 0.29 eV (using 10-fold cross-validation). Table S1 in the Supporting Information compares the performance of the RF model with MLP, SVM, and KRR models which are all found to perform much worse than the RF model. 

Fig. \ref{fig3} (b) shows the most important features and their \% importance as computed using the Gini importance method. The most important features include properties like dielectric constants, effective masses of electrons and holes, and atomic packing fractions that are also considered in well-known physical theories of EBE. For instance, the average dielectric constant and the hole-effective mass features are also included in the Wannier-Mott (WM) model. In the WM model, the $\mathrm{EBE}_\mathrm{WM}=\frac {\mu e^4} {2 \hbar^2 \epsilon^2}$, where  $\mu=\frac {\mu_e \mu_h} {\mu_e+\mu_h}$ is the reduced effective mass of an electron and a hole, $e$ is the charge of an electron, $\hbar$ is the reduced Planck's constant and $\epsilon$ is the dielectric constant of the material. $\mathrm{EBE}_\mathrm{WM}$, can thus be obtained from ground state properties without explicit BSE simulations. 

In the case of Wannier-Mott (WM) excitons, the Coulomb attraction between $e$-$h$ pairs is screened to a larger extent resulting in exciton wavefunction that is spread over multiple unit cells and low EBE. Since a majority of the materials in our dataset have a low ($<$ 1 eV) EBE, the Wannier-Mott model is applicable to them. 

\begin{figure}
\centering
\includegraphics[width=0.5\linewidth]{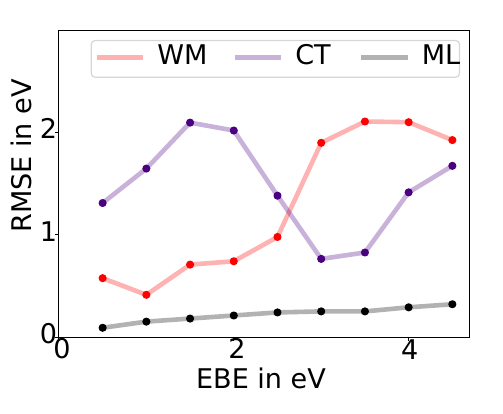}
\caption{The RMSE accuracy of both Wannier-Mott (WM) and Charge transfer/Frenkel (CT) models as well the ML model as a function of the EBE. The RMSE as a function of EBE has been calculated by considering only materials with EBE in a 1 eV window around a certain EBE value. Our results show ML model performs better in any energy window than WM or CT model. WM model works fairly well in the case of low EBE but fails in the high EBE region. CT model is less accurate than WM model in low EBE region but works better in the high EBE region.} 
\label{fig4}
\end{figure}

In Fig. \ref{fig4} we compare the RMSE accuracy of both the WM model and the ML model as a function of the EBE. This RMSE as a function of EBE has been calculated by considering only materials with EBE in a 1 eV window around a certain EBE value. Our results show that the ML model has a much lower RMSE than the WM model. Furthermore, while consistently poorer than the ML model, the WM works comparatively well at low EBE but fails dramatically in the high EBE region. This is not surprising, since the materials that have very high EBE in the range of 4-5 eV are expected to exhibit Frenkel or Charge Transfer (CT) type excitons. CT excitons are more localized with very strong Coulomb attraction between $e$-$h$ pairs and therefore have high EBE. The EBE of a CT exciton is given by the expression, $\mathrm{EBE}_\mathrm{CT}=\frac {e^2} {4 \pi \epsilon r_{CT}} $ where $r_{CT}$ is the separation between the electron and hole of an exciton or radius of exciton wavefunction. Unlike the WM model, this model can not be used to predict the EBE of solid-state materials using ground-state DFT computed properties, as $r_{CT}$ can not be computed without solving the BSE. CT excitons are usually localized in a length scale of the order of the size of a unit cell of materials and are also expected to have smaller $r_{CT}$ for materials with tighter packing efficiency. Thus one can assume that $r_{CT} = fV^{-1/3}$, where $f$ is a dimensionless proportionality constant and V is the volume of the unit cell, allowing the estimation $\mathrm{EBE}_\mathrm{CT}$ from ground state properties without the need of BSE simulations.

In Fig. \ref{fig4} we present the RMSE accuracy of EBE obtained from the CT model (f=0.5) as a function of the EBE. In comparison to the ML model, the CT model is consistently poorer with high RMSE values. However, as expected, it performs better than the WM model in the high EBE region. 

Overall, the ML model performs much better in any energy window in comparison to the WM or CT model. We think this superior predicting capability comes from the inclusion of additional material properties not present in the WM model such as packing fraction and range of electron and hole effective masses and dielectric constants. By including such attributes our ML model is capturing the physics of not only the low EBE excitons but also the higher EBE regime where the CT model is more perhaps applicable than the WM model. Therefore, one can in principle build a more general empirical model for excitons based on the properties revealed by our ML model.   



\subsection{ML Classification of Exciton Binding Energies and Absorption-Related Properties}
\label{classification}

\begin{figure}
\centering
\includegraphics[width=\linewidth]{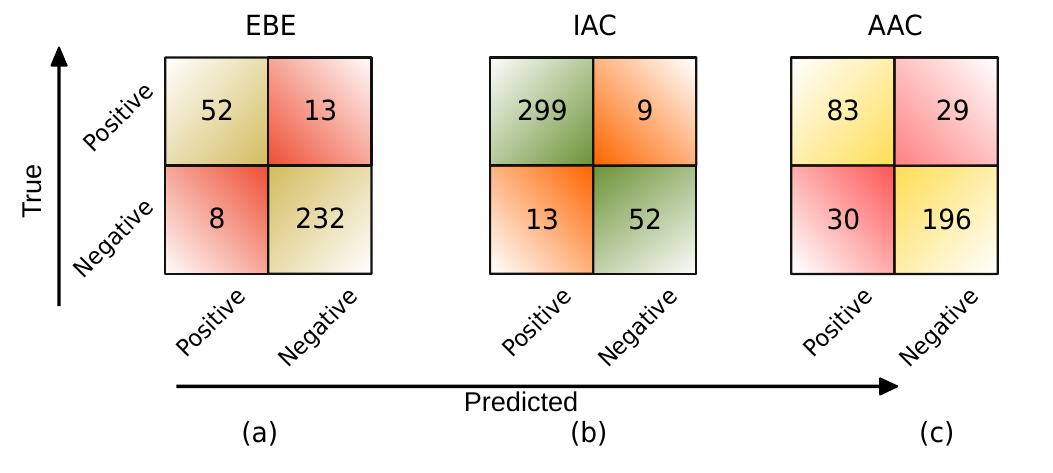}
\caption{Confusion matrix illustrating the number of true positives, (TP), true negatives (TN), false negatives (FN), and false negatives (FN), for predicting if materials possess a low exciton binding energy (EBE), $<$ 0.2 eV, high integrated absorption coefficient (IAC), $>$ 10.5 $\times$10$^4$ cm$^{-1}$ eV, and anisotropy in absorption coefficient, $>$ 0.8. All the incorrect classifications (TPs and TNs) are shown in different shades of red and correct classifications (FPs and FNs) are shown in shades of green and and yellow.} 
\label{fig5}
\end{figure}
In the previous two sections, we discussed the capabilities of our ML model in predicting accurate QPG and EBE of a wide variety of materials by solving a regression problem.  This undoubtedly has a lot of potential applications in identifying materials with specific bandgap requirements such as for ultra-wide bandgap semiconductors for power electronics applications or optical devices in visible or UV light applications. However, from a material designing/discovery point of view, an equally useful exercise would be to identify materials with desirable optical/excitonic properties such as having a low EBE, which is a classification problem. In this section, we apply classification algorithms to classify materials based on their excitonic properties.

Among excitonic properties, low EBEs are preferred in applications where free $e$-$h$ pairs are desired for example in photocatalytic materials. \cite{biswas2021excitonic},  In addition, two other parameters derived from BSE obtained absorption spectra—the integrated absorption coefficient (IAC) in the solar wavelength range of interest and anisotropy in absorption coefficient (AAC)—are useful to quantify the potential of a material for solar-energy absorption, for example in photovoltaics and photocatalysts. The methods section describes the calculation of IAC and AAC from the frequency-dependent absorption spectra. In a previous study, we have established that low EBE materials are those that have EBE smaller than 0.2 eV, high (visible/UV)-light IAC materials have an IAC larger than $>$ 10.5 $\times$10$^4$ cm$^{-1}$ eV, and high AAC materials have AAC $\geq$ 0.8. 

Fig. \ref{fig5} shows the results of classification obtained by the RF method in the form of confusion matrixes. Fig. \ref{fig5} (a) shows the confusion matrix for classifying materials as low EBE. Fig. \ref{fig5} (b) shows the matrix for classifying materials as high UV-light IAC and Fig. \ref{fig5} (c) for classifying materials as as high AAC materials. For the EBE classification, only 21 of the 305 materials were classified incorrectly resulting in a high classification accuracy of 93~$\%$. The classification models for the IAC and AAC resulted in an accuracy of 94~$\%$ and 83~$\%$ for the 355 materials in the dataset, respectively. Additionally, we employed the AdaBoost, stochastic gradient descent (SGD), and MLP \cite{bonaccorso2017machine, mohammed2016machine} methods for the classification. However, the RF performed best among the four methods. Supporting Information Table S2 shows the comparison between the four methods. 

 The feature set used for IAC and AAC classification was selected following a similar strategy employed for the QP gap and EBE prediction described earlier. In the supporting information figure S2 (IAC) and S3 (AAC) we have shown the most important features along with their \% importance. In the case of IAC prediction mean dielectric constant (67.1~$\%$) and DFT computed bandgap ($4.6~\%$) emerge as the two most important features. The emergence of these two properties as the most important features can be understood from the fact that for a high absorption in the visible spectrum, 1.7-3.5 eV, one needs to have a material with a QP gap in the same range and also needs to have significant absorption coefficient ($\propto \epsilon(\omega)$ ) in that energy range. As we have seen in Fig. \ref{fig2} for the QP gap prediction mean dielectric constant and DFT computed bandgap are the two most important features, it is not surprising that they are also equally important for the IAC predictions. Moreover, the high importance of the mean dielectric constant also signifies a high degree of correlation between the static dielectric constant of a material and frequency-dependent dielectric function. In the prediction of anisotropy in visible light absorption (AAC) we find that the range of dielectric constant ($ \max \{\epsilon_x,\epsilon_y,\epsilon_z\} - \min \{\epsilon_x,\epsilon_y,\epsilon_z\}$) is the most important feature (58.8~$\%$). All the other important features in the prediction of AAC have importance $< 5~\%$. Therefore, one can identify a material with a high degree of anisotropy in visible light absorption by looking at the anisotropy in the static dielectric constant, which once again highlights the importance of static dielectric constants in the excitonic properties. The Materials Project (MP)\cite{jain2013commentary} database currently holds $\sim$ 150,000 materials, but only $\sim$ 7000 (4.7$~\%$) of them have computed static dielectric constants. We believe that the static dielectric constant using DFT is quite inexpensive to calculate but is a crucial parameter to understand a material applicability for a wide variety of electronic and optoelectronic applications and therefore it would be useful to compute and curate it for more materials in existing materials databases.  

\subsection{Screening materials from MP database}
\label{screening}
              
\begin{figure} 
\centering
\includegraphics[width=\linewidth]{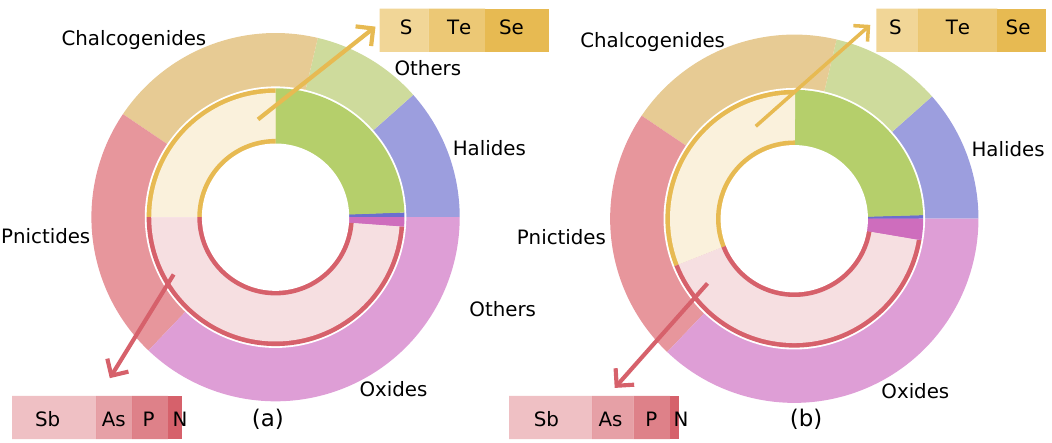}
\caption{Materials obtained from the MP database and selected for visible (a) and UV light (b) applications through the ML-based screening process are categorized into different chemical compositions. The outer rings represent the fraction of the starting set of $\sim$ 7000 materials belonging to different chemical groups whereas the inner rings represent the same for selected (159 for visible and 203 for UV) materials. In the case of visible light absorption, we see significantly fewer oxides pass through the screening process, whereas, pnictides and chalcogenides comprise a major fraction of the selected materials.} 
\label{fig6}
\end{figure}

The high accuracy ($~90~\%$) of our ML classifiers enables their application to a larger material set to screen for materials suitable for photoabsorption-based applications such as photovoltaics, and photocatalysis. 

We perform such a screening on $\sim$ 7083 materials that have static dielectric constants and DFT computed band structure available in the MP database. We find that out of 7000 materials, only 159 passed the criterion of low EBE, high IAC, and high AAC in the visible-light region, 1.7 -3.5 eV. In the case of the UV-light absorption, 3.5-4.2 eV, we found 237 materials passed the aforementioned criteria. 

Fig. \ref{fig6} examines the chemical compositions of the materials in the starting set of $\sim$ 7000 materials (outer rings) as well as ones that emerged as suitable for visible-light applications, \ref{fig6} (a)), and  UV-light, Fig. \ref{fig6} (b), respectively, inner rings. We find that most of the visible-light materials are either pnictides or chalcogenides. Note that almost half of the starting set of materials, 47~$\%$, are oxides but only 2 oxides pass through the screening. This is not surprising since oxides mostly have larger bandgaps than the visible light region. More oxides, six, are found in the screened materials for the UV light region. Most of the materials for UV absorption also belong to pnictides or chalcogenides. Furthermore, we find that almost half of the selected chalcogenides are tellurides. In the supporting information tables, S6 and S7 we list all the screened materials for both visible and UV absorption respectively along with their ML-predicted QP gap and EBE values. In Tables S6 and S7, we have also reported whether a screened material has been already synthesized before and has an ICSD ID and the computed value of energy above hull. We find that the majority of these materials 193 out of 234 materials have been already synthesized. Moreover, we find that 168 out of 193 aforementioned materials have computed energy above the hull value of 0 eV. An examination of the screened materials shows that several of these materials have been studied in the context of photoabsorption-related applications, for example, GeTe \cite{noman2018germanium}, AlSb \cite{dhakal2011alsb}, SnSe \cite{minnam2016perspectives}, etc. Thus it is likely that the other screened materials can be promising novel materials for photoabsorption-related applications. 



\section{Conclusion}
In this study, we report the largest dataset of excited state properties of bulk materials calculated using the state-of-the-art GW-BSE formalism. This database has been created using the open-source Python workflow package $py$GWBSE and made publicly available through the website \href{hydrogen.cmd.lab.asu.edu/gwbse-data}{https://hydrogen.cmd.lab.asu.edu/gwbse-data}. Using this dataset we have developed ML models that can predict the QP gap and EBE of a diverse set of materials with excellent accuracy. Despite having a limited dataset size of $\sim$ 350 materials, the ML models predict the QP gap with an accuracy of 0.36 eV (RMSE) and EBE with an accuracy of 0.29 eV. Moreover, we demonstrated that using the ML models developed on this dataset we can screen materials based on excited state properties such as EBE, IAC, and AAC by using only ground state DFT computed properties. Using our ML models we utilize the relevant existing DFT computed date in the MP database, available for 7000 materials, and screen 159 materials for visible and 203 materials for UV light-based applications. This work presents a robust framework to utilize the underlying potential of the large datasets of ground-state properties available in open-source materials databases, allowing us to obtain excited state properties without the need for explicit simulations.

\section{Computational Methods}
\label{methods}

\subsection{GW-BSE calculations}
\label{gw-bse-methods}

Quasi-particle (QP) energies, the energy required to add or remove an electron from an interacting many-electron system is not a ground state property of the system and therefore can't be computed accurately using DFT. These QP energies can be computed correctly at a reasonable computational cost using many-body perturbation theory within GW approximation. Within GW approximation one computes the self-energy as the product of one-particle Green's function (G) and the screened Coulomb interaction (W) \cite{hybertsen1986electron, onida2002electronic}. The QP energies are evaluated as a correction to the KS eigenvalues using first-order perturbation by assuming the difference between exchange-correlation potential ($V_{xc}$) and self-energy is sufficiently small. QP energies, and thus QPGs, of all the materials in this study were computed using the one-shot G$_0$W$_0$ method via the high-throughput workflow code $py$GWBSE \cite{biswas2023py}. $py$GWBSE workflows allow high-throughput first principles atomistic simulations via the VASP\cite{shishkin2006implementation} package by high-throughput automated input file generation, submission to supercomputing platforms, analysis of post-simulation data, and storage of metadata and data in a MongoDB database. A plane wave cutoff of 500 eV and a $k$-grid with reciprocal density 200 $\AA^{-3}$ was used for all the GW-BSE calculations. A fixed value of 100 eV for the screened Coulomb energy cutoff and 80 for the number of frequency grid points was used for all the GW calculations since these parameters don't display a significant dependence on the material of choice.\cite{biswas2023py}  The number of unoccupied bands for the GW simulations was selected such that the QPGs converged within $<$0.1 eV. Once the QP energies were obtained, the absorption spectra, $\epsilon(\omega)$, and EBE were computed by solving the Bethe-Salpeter equation (BSE). BSE is a two-particle equation that explicitly includes the electron-hole interactions or the excitonic effects within the Tamn-Dancoff approximation \cite{rohlfing2000electron, sander2015beyond}. The number of valence ($v$) and conduction ($c$) bands included in the BSE calculation are selected such that the vertical $v$ $\rightarrow$ $c$ transitions of energy less than 3 eV were obtained. From a BSE solution, the optical gap is the lowest energy excitation. The EBEs are the difference between the optical gaps and the direct QP gaps.

To quantify the fraction of incident light that can be absorbed by a material in a desired frequency range, we can compute the integrated absorption coefficient, IAC. IAC is obtained by integrating the BSE computed frequency-dependent absorption coefficient, $\alpha_\mathrm{int}$.\cite{biswas2023py} In the case of light polarization along $x$ axis,

\begin{multline}
\alpha_\mathrm{int}^x = \int_{\omega_{min}}^{\omega_{max}}{\alpha^x(\omega)} d\omega \\  
\mathrm{where} \hspace{8pt}
	\alpha^x(\omega)=\frac{2 \pi (\lvert \sqrt{(\epsilon^x_1(\omega))^2+ (\epsilon^x_2(\omega))^2} \rvert - \epsilon^x_1(\omega))^{1/2}} {\lambda} 
\end{multline}

where $\epsilon_1(\omega)$ is the real and $\epsilon_2(\omega)$ the imaginary part of the dielectric function. $\omega$ and $\lambda$ are the frequency and wavelength of incident radiation. The $\epsilon_2(\omega)$ is obtained by solving BSE and $\epsilon_1(\omega)$ is computed using the Kramers–Kronig relation.  


To assess whether a material has a preference for absorbing light of certain polarization we calculate the anisotropy in absorption coefficient (AAC), $\alpha^\mathrm{aniso}_\mathrm{int}$. 
$\alpha^\mathrm{aniso}_\mathrm{int}$ is defined as the ratio of $min (\alpha^{x}_\mathrm{int}, \alpha^{y}_\mathrm{int}, \alpha^{z}_\mathrm{int})$ and $max(\alpha^{x}_\mathrm{int}, \alpha^{y}_\mathrm{int}, \alpha^{z}_\mathrm{int})$.

\subsection{Machine Learning}
\label{ml-methods}

All the ML models were created using the Scikit-Learn ML library.\cite{kramer2016scikit} Over 150 features were used for training the ML models for this work. Section II in the supporting information describes the features considered in this work. The features include DFT-computed material properties that are available in the MP database including the DFT bandgaps and the dielectric constants. In order to include the effective masses as features, we computed the effective masses from the bandstructures available in the MP database by employing the sumo \cite{ganose2018sumo} code and the MP Application Programming Interface (API). 

The most important features were determined by computing the Gini Importance \cite{breiman2017classification} method. The Gini Importance of each feature is calculated as the decrease in node impurity weighted by the probability of reaching that node. Furthermore, for the prediction of QPG and EBE, we chose the minimum number of features that were needed to obtain an RMSE value converged within 0.01 eV.  





\section*{Acknowledgements}
This research was supported by the U.S. Department of Energy (DOE), Office of Science, Basic Energy Sciences (BES), under Award Number DE-SC0024184 (machine learning), ULTRA, an Energy Frontier Research Center funded by the U.S. Department of Energy (DOE), Office of Science, Basic Energy Sciences (BES), under Award $\#$ DE-SC0021230 (GW-BSE high-throughput simulations), Arizona State University start-up funds, and by the National Science Foundation (NSF) under Award Number 2235447 (ab initio simulations). The authors acknowledge the San Diego Supercomputer Center under the NSF-XSEDE and NSF-ACCESS Award No. DMR150006, and the Research Computing at Arizona State University for providing HPC resources. This research used
resources from the National Energy Research Scientific Computing Center, a DOE Office of Science User Facility supported by the Office of Science of the U.S. Department of Energy under Contract No. DE-AC02-05CH11231. 

\section*{Conflict of interest}
The authors declare no conflicts of interest.

\section*{Supporting Information}
The supporting information contains the performance of various ML algorithms applied in this study, a detailed description of the features used in ML models, and a list of materials shortlisted for visible and UV light-based applications along with their ML-predicted properties.  

\bibliography{article}

\end{document}